# Superconductivity in the $Cu(Ir_{1-x}Pt_x)_2Se_4$ Spinel

Huixia Luo[1], Tomasz Klimczuk[2,3], Lukas Müchler[4], Leslie Schoop[1,5], Daigorou Hirai[1], M.K. Fuccillo[1], C. Felser[4] and R. J. Cava[1*]

[1]Department of Chemistry, Princeton University, Princeton, New Jersey 08544, USA

[2] Faculty of Applied Physics and Mathematics, Gdansk University of Technology, Narutowicza 11/12, 80-233 Gdansk, Poland

[3] Institute of Physics, Pomeranian University, Arciszewskiego, 76-200 Slupsk, Poland

[4] Max-Planck-Institut für Chemische Physik Fester Stoffe, 01187, Dresden, Germany

[5] Graduate School Material Science in Mainz, 55099, Mainz, Germany

**ABSTRACT**

We report the observation of superconductivity in the $CuIr_2Se_4$ spinel induced by partial substitution of Pt for Ir. The optimal doping level for superconductivity in $Cu(Ir_{1-x}Pt_x)_2Se_4$ is x = 0.2, where $T_c$ is 1.76 K. A superconducting $T_c$ vs. composition dome is established between the metallic, normal conductor $CuIr_2Se_4$ and semiconducting $CuIrPtSe_4$. Electronic structure calculations show that the optimal $T_c$ occurs near the electron count of a large peak in the calculated electronic density of states and that $CuIrPtSe_4$ is a band-filled insulator. Characterization of the superconducting state in this heavy metal spinel through determination of $\Delta C/\gamma T_c$, indicates that it is BCS-like. The relatively high upper critical field at the optimal superconducting composition ($H_{c2}(0)$ = 3.2 T) is much larger than that reported for analogous rhodium spinels and is comparable to or exceeds the Pauli field ($\mu_0 H_p$), suggesting that strong spin orbit coupling may influence the superconducting state. Further, comparison to doped $CuIr_2S_4$ suggests that superconductivity in iridium spinels is not necessarily associated with the destabilization of a charge-ordered spin-paired state through doping.

* Corresponding author: R. J. Cava. E-mail address: rcava@princeton.edu

# 1. Introduction

Materials with the spinel crystal structure display a wide range of structural, magnetic, and electronic properties, but are rarely superconducting. The oxide spinel $LiTi_2O_4$ displays the highest $T_c$ in this structural family [1,2], and, to the best of our knowledge, the only other spinel superconductors found are ternary metal chalcogenides such as $CuRh_2S_4$, $CuRh_2Se_4$ and electron doped $CuIr_2S_4$ [3-6]. The heavy metal chalcogenide spinels have been of particular interest due to the presence of a metal-insulator (M-I) transition on cooling or under pressure [7-10]. The temperature-induced M-I transition in $CuIr_2S_4$ at $T \approx 230$ K, for example, is accompanied by a complex structural transition that concurrently creates both charge ordering and metal-metal pairing [11,12]. On Zn substitution for Cu in the $Cu_{1-x}Zn_xIr_2S_4$ solid solution, the M-I transition is suppressed and superconductivity appears, with a maximum $T_c$ of 3.4 K near $x = 0.3$ [6]. It is natural to associate the appearance of superconductivity in this system with the doping-induced destabilization of the charge-ordered, spin-paired state, in analogy to what is found for the suppression of the charge density wave (CDW) in doped chalcogenides [13], the charge disproportionation in $BaBiO_3$ [14], and the magnetism in the cuprates and pnictides [15,16].

$CuIr_2Se_4$ possesses the same spinel structure as $CuIr_2S_4$ (inset, Figure 1a [17]), but does not undergo a structural phase transition on cooling. Also, it maintains its metallic conduction from room temperature down to 0.5 K; no M-I or superconducting transitions are observed at ambient pressure [18-20]. Under pressure, however, a M-I transition is found above 2.8 GPa [8,21]. Thus, though no charge-ordering or metal-metal pairing has been observed for $CuIr_2Se_4$, it appears to be at the borderline of such behavior [22], i.e. it may have an incipient tendency toward such instabilities. Because on first sight it appears to be an ordinary metal, a smaller number of studies have been performed on $CuIr_2Se_4$. The complex behavior of $CuIr_2S_4$, the expected strong spin orbit coupling of $5d$ Ir, and the geometric frustration intrinsic to the spinel structure suggest on the other hand that the properties of $CuIr_2Se_4$ should be considered more carefully. Here we show that superconductivity can be induced through the appropriate chemical substitution.

We report the synthesis and physical properties of the spinel solid solution $Cu(Ir_{1-x}Pt_x)_2Se_4$ ($0 \leq x \leq 0.5$), characterized via X-ray diffraction (XRD), magnetization, resistivity and heat capacity measurements. Superconductivity is observed for $Cu(Ir_{1-x}Pt_x)_2Se_4$ ($0.1 \leq x \leq 0.35$) with a maximum $T_c$ = 1.76 K for $Cu(Ir_{0.8}Pt_{0.2})_2Se_4$. The superconductivity occurs between the compositions of the metallic host compound $CuIr_2Se_4$ and semiconducting $CuIrPtSe_4$ (x = 0.5), which we show by electronic structure calculations to be a conventional band-filling semiconductor; these calculations also show a peak in the expected electronic density of states (DOS) at the Fermi Energy ($E_F$) near the composition where $T_c$ is optimized. Comparison to the superconducting doped $CuIr_2S_4$ system suggests that superconductivity in iridate spinels is not necessarily associated with the destabilization of a charge-ordered spin-paired state through doping. Finally, the high $H_{C2}(0)$ we observe for $Cu(Ir_{0.8}Pt_{0.2})_2Se_4$ relative to the superconducting rhodium-based chalcogenide spinels suggests that strong spin orbit coupling may influence the superconducting state in this material.

**2.1 Experimental**

Polycrystalline samples of $Cu(Ir_{1-x}Pt_x)_2Se_4$ were synthesized by conventional solid state reaction. Mixtures of high-purity fine powders of Cu (99.5%), Ir (99.95%), Pt (99.95%), and Se (99.999%) in the appropriate stoichiometric ratios were thoroughly ground, pelletized and heated in sealed quartz tubes at 850 °C for 96 h. Subsequently, the as-prepared powders were reground, re-pelletized and sintered again at 1123 K for 48 h. Samples with higher Pt content required several cycles of heating and grinding. Powder X-ray diffraction (PXRD, Bruker D8 Focus, Cu Kα radiation, graphite diffracted beam monochromator) was used to structurally characterize the samples. Measurements of the temperature dependence of the electrical resistivity, heat capacity and magnetization were performed in a Quantum Design Physical Property Measurement System (PPMS) from 2 to 300 K. Selected resistivities, for $Cu(Ir_{1-x}Pt_x)_2Se_4$ (x = 0.1, 0.2, 0.3, 0.35), and heat capacities, for $Cu(Ir_{0.8}Pt_{0.2})_2Se_4$ and $Cu(Ir_{0.7}Pt_{0.3})_2Se_4$, were measured in the PPMS equipped with a $^3$He cryostat. Seebeck coefficient ($S$) measurements were performed using a modified MMR Technologies SB-100 Seebeck measurement system.

**2.2 Calculational Details**

The calculations were performed in the framework of density functional theory (DFT) using the WIEN2K code with a full-potential linearized augmented plane-wave and local orbitals [FP-LAPW + lo] basis [23] together with the Perdew Becke Ernzerhof (PBE) parametrization [24] of the Generalized Gradient Approximation (GGA) as the exchange-correlation functional with spin orbit coupling (SOC) and no spin polarization. The plane-wave cutoff parameter $RK_{MAX}$ was set to 8 and the Brillouin zone was sampled by 10,000 k points. To simulate the doping, the virtual crystal approximation (VCA) was employed. Experimental lattice constants were used and the free internal parameters were optimized by minimizing the forces.

## 3. Results and Discussion

The room-temperature lattice parameters for polycrystalline $Cu(Ir_{1-x}Pt_x)_2Se_4$ are displayed in Figure 1(a). Figure 1(b) shows the powder X-ray diffraction patterns for the $Cu(Ir_{1-x}Pt_x)_2Se_4$ samples. The spinel phase (*Fd-3m*, #227) is found for $0 \leq x \leq 0.5$, though very small amounts (< 5%) of $IrSe_2$ are found in some preparations. With increasing Pt content x in $Cu(Ir_{1-x}Pt_x)_2Se_4$, the lattice parameter (*a*) increases linearly from 10.3199(3) Å (x = 0) to 10.3864(2) Å (x = 0.5, ($CuIrPtSe_4$)), the limit of the solid solution for our synthesis conditions, consistent with Vegard's law. The relative change, $\Delta a/a$, with increasing x from 0 to 0.5 in $Cu(Ir_{1-x}Pt_x)_2Se_4$ is about 0.6 %.

The full measured temperature range of electrical resistivity, $\rho(T)$, for x = 0.2 and x = 0.3, is shown in the main panel of Figure 2. The samples show a metallic temperature dependence ($d\rho/dT > 0$) in the temperature region of 2 to 300 K, similar to the $CuIr_2Se_4$ host compound [19,20]. For all the metallic Pt-doped samples the residual-resistivity parameter is small, RRR= $\rho_{300K}/\rho_n$ < 1.6, reflecting the presence of substantial atomic disorder. The disorder effect increases with increased Pt doping, as is revealed by an increase of the residual resistivity, $\rho_n$, with increasing Pt concentration. A sharp drop of $\rho(T)$ is seen at low temperatures signifying the onset of superconductivity. The temperature dependence of the resistivity in the vicinity of the superconducting transition is shown in the inset of Figure 2. A very sharp transition, with $\Delta T_c$ < 0.1 K, is observed for $Cu(Ir_{0.8}Pt_{0.2})_2Se_4$, the sample with the highest observed $T_c$ = 1.76 K. ($T_c$ is taken as the intersection of the extrapolation of the steepest slope of the resistivity $\rho(T)$ in the superconducting transition region and the extrapolation of the normal state resistivity ($\rho_n$) [25]).

The superconducting transition for the two optimal superconducting samples (x = 0.2 and x = 0.3) was further examined through temperature dependent measurements of the electrical resistivity under applied magnetic field. Figure 3 presents $\rho(T,H)$ obtained for $Cu(Ir_{0.8}Pt_{0.2})_2Se_4$ and $Cu(Ir_{0.7}Pt_{0.3})_2Se_4$. Based on the $T_c$ determined for different magnetic fields, the upper critical field values, $\mu_0H_{c2}$, are plotted vs. temperature in the insets to Figure 3. A clear linear dependence of $\mu_0H_{c2}$ vs. T is seen; the solid line through the data shows the best linear fit with the initial slope $dH_{c2}/dT$ = -2.6 T/K for $Cu(Ir_{0.8}Pt_{0.2})_2Se_4$. Similarly, for $Cu(Ir_{0.7}Pt_{0.3})_2Se_4$, the slope obtained is $dH_{c2}/dT$ = -3.2 T/K. By using the Werthamer-Helfand-Hohenberg (WHH) expression, $\mu_0H_{c2}(0) = -0.693T_c (dH_{c2}/dT_c)$ [26], we estimate the zero temperature upper critical fields as $\mu_0H_{c2}(0)$ = 3.2 T and 3.6 T, for $Cu(Ir_{0.8}Pt_{0.2})_2Se_4$ and $Cu(Ir_{0.7}Pt_{0.3})_2Se_4$, respectively. From $\mu_0H_{c2} = \frac{\phi_0}{2\pi\xi_{GL}^2}$, where $\phi_o$ is the quantum of flux, the Ginzburg-Laudau coherence length can be estimated as $\xi_{GL}(0)$ = 101 Å and 96 Å for $Cu(Ir_{0.8}Pt_{0.2})_2Se_4$ and $Cu(Ir_{0.7}Pt_{0.3})_2Se_4$, respectively. The values of upper critical fields $\mu_0H_{c2}(0)$ obtained for the $Cu(Ir,Pt)_2Se_4$ materials are significantly larger than those reported for the higher $T_c$ spinels $CuRh_2Se_4$, ($T_c$ = 3.48 K, $\mu_0H_{c2}(0)$ = 0.44 T) and $CuRh_2S_4$ ($T_c$ = 4.7K, $\mu_0H_{c2}(0)$ = 2.0 T), [5]. Assuming a Lande g-factor of 2, the measured $\mu_0H_{c2}(0)$ is comparable (x = 0.2) or higher (x = 0.3) than the weak-coupling Pauli field $\mu_0H_P$ = 1.84$T_c$ = 3.2 T and 3.0 T for $Cu(Ir_{0.8}Pt_{0.2})_2Se_4$ and $Cu(Ir_{0.7}Pt_{0.3})_2Se_4$, respectively. This comparison suggests that strong spin orbit coupling may play a role in the characteristics of the superconducting state in these materials.

For higher Pt concentrations, the superconductivity disappears. Semiconducting behavior ($d\rho/dT < 0$) is observed for x = 0.5 ($CuIrPtSe_4$), as is shown in Figure 4 (a). The resistivity does not obey a simple activated temperature dependence ($\rho = \rho_0\exp(-\Delta/T)$) for any part of the measured temperature range (Figure 4 (b)). At low temperatures, the temperature dependence of the resistivity is consistent with expectations for three-dimensional variable range hopping, where $\rho = \rho_0\exp(-T_0/T)^{1/n}$, and n = 4 [27] (inset Figure 4(b)). This is consistent with our overall conclusion that $CuIrPtSe_4$, with a random

Pt/Ir distribution on the octahedral sites, is a strongly disordered low density of states (see heat capacity data, below) semiconductor, although further transport study on single crystals would be necessary to establish that state conclusively. Seebeck coefficient measurements (Inset, Figure 4(a)) show that $CuIrPtSe_4$ is *p*-type and has a relatively large Seebeck coefficient near room temperature, characteristic of semiconducting materials with carriers near the top of the valence band, consistent with electronic structure calculations (see below).

The electronic heat capacity data for $Cu(Ir_{0.8}Pt_{0.2})_2Se_4$ in the vicinity of $T_c$ are presented in Figure 5. The main panel shows the temperature dependence of the zero-field electronic specific heat $C_{el}/T$. The good quality of the sample and the bulk nature of the superconductivity are supported by the presence of a sharp anomaly at $T_c = 1.76$ K, in excellent agreement with the $T_c$ determined by $\rho(T)$. From the specific heat measured in zero magnetic field, we estimate $\Delta C/T_c = 26.07$ mJ·mol$^{-1}$·K$^{-2}$ for $Cu(Ir_{0.8}Pt_{0.2})_2Se_4$ and $\Delta C/T_c = 22.61$ mJ·mol$^{-1}$·K$^{-2}$ for $Cu(Ir_{0.7}Pt_{0.3})_2Se_4$ (data not shown), respectively. The low temperature heat capacity obeys the relation of $C_p = \gamma T + \beta T^3$ (inset Figure 5), where $\gamma$ and $\beta$ describe the electronic contribution and the phonon contribution to the heat capacity, respectively, the latter of which is a measure of the Debye Temperature ($\Theta_D$). We fitted $C_p(T)/T$ vs. $T^2$ with $C_p(T)/T = \gamma + \beta T^2$ in the temperature range of 2 K - 7 K, which yields the electronic specific heat coefficient $\gamma = 16.5$ mJ·mol$^{-1}$·K$^{-2}$ and phonon specific heat coefficient $\beta = 1.41$ mJ·mol$^{-1}$·K$^{-4}$ for $Cu(Ir_{0.8}Pt_{0.2})_2Se_4$. Using the value of $\beta$, we estimate the Debye temperature by the relation $\Theta_D = (12\pi^4 nR/5\beta)^{1/3}$, where n is the number of atoms per formula unit (n = 7), and R is the gas constant; $\Theta_D = 212$ K for $Cu(Ir_{0.8}Pt_{0.2})_2Se_4$. The normalized specific heat jump values $\Delta C/\gamma T_c$ for $Cu(Ir_{0.8}Pt_{0.2})_2Se_4$ and $Cu(Ir_{0.7}Pt_{0.3})_2Se_4$ (data not shown) are found to be 1.58 and 1.44, respectively, which are near that expected for the Bardeen-Cooper-Schrieffer (BCS) weak-coupling value (1.43), confirming bulk superconductivity. Using the Debye temperature $\Theta_D$, critical temperature $T_c$, and assuming $\mu^* = 0.15$, the electron-phonon coupling constant ($\lambda_{ep}$) can be calculated from the inverted McMillan formula [28]:

$$\lambda_{ep} = \frac{1.04 + \mu^* \ln\left(\frac{\theta_D}{1.45T_C}\right)}{(1 - 0.62\mu^*)\ln\left(\frac{\theta_D}{1.45T_C}\right) - 1.04}.$$

The values of $\lambda_{ep}$ obtained range from 0.51 to 0.57 for x = 0.1 (underdoped) and x = 0.2 (optimally doped) compositions, respectively, and suggest weak coupling superconductivity. With the Sommerfeld parameter ($\gamma$) and the electron-phonon coupling ($\lambda_{ep}$), the density of states at the Fermi level can be calculated from: $N(E_F) = \frac{3}{\pi^2 k_B^2 (1 + \lambda_{ep})} \gamma$. The highest N(E$_F$) = 4.45 states/eV f.u. was obtained for optimally doped Cu(Ir$_{0.8}$Pt$_{0.2}$)$_2$Se$_4$. The measured and calculated properties of the materials are summarized in Table 1.

Figure 6 summarizes our general electronic characterization of the Cu(Ir$_{1-x}$Pt$_x$)$_2$Se$_4$ phase. Figure 6a shows the measured electronic contribution to the specific heat. The Sommerfeld parameter ($\gamma$) is relatively small (7.0 mJ mol$^{-1}$ (formula) K$^{-2}$) for CuIr$_2$Se$_4$, and increases to 16.5 mJ mol$^{-1}$ (formula) K$^{-2}$ at the composition where Tc is highest (x = 0.2) before it decreases to a significantly smaller value (3.0 mJ mol$^{-1}$ (formula) K$^{-2}$) for the semiconducting material CuIrPtSe$_4$. The Debye temperature obtained from the fits shows some noise but we interpret the data to show that it does not change much over the composition range of the solid solution. This is expected because the lattice parameter changes only by 0.6 % and the molar mass of the compound varies relatively little (0.4 %) over the range of the solid solution. Finally, Figure 6c summarizes the experimental results on an electronic phase diagram for Cu(Ir$_{1-x}$Pt$_x$)$_2$Se$_4$ (0 ≤ x ≤ 0.5). With Pt doping, the superconducting transition appears for x ≥ 0.1. The maximum T$_c$ of around 1.76 K is found for Cu(Ir$_{0.8}$Pt$_{0.2}$)$_2$Se$_4$. T$_c$ then disappears for x > 0.35. The sample with x = 0.5 shows semiconductor behavior.

Figure 7 shows the calculated density of states, the Fermi surfaces, and the band structure in the vicinity of E$_F$ for Cu(Ir$_{0.7}$Pt$_{0.3}$)$_2$Se$_4$. Within the VCA, Cu(Ir$_{1-x}$Pt$_x$)$_2$Se$_4$ is a metal for 0 ≤ x < 0.5, and a conventional band insulator for x = 0.5. Upon doping electrons, the calculated density of states at the Fermi level rises quickly and reaches a maximum for x = 0.3. This maximum in the DOS arises from a van Hove singularity (vHs) in the band structure at the Γ-point and is quite narrow in energy. The bands at the Fermi level in $Cu(Ir_{0.7}Pt_{0.3})_2Se_4$ consist mostly of Se $p$- and Ir/Pt $d$-states with little intermixing of Cu states. Bands with mostly Se-p character give rise to the vHs, with little hybridization from Ir/Pt $d$-states. This is an indication of the important role of the anions in the electronic properties and therefore the superconductivity in this spinel. The role of vHs for superconductivity has been pointed out in numerous scenarios [29-32], although it is generally expected to have the most influence on superconductivity in 2D electronic systems, which is not the case here. In any case the maximum observed $T_c$ for $Cu(Ir_{1-x}Pt_x)_2Se_4$ is near the electron filling of a calculated peak in the DOS due to the vHs, implying that the superconductivity arises in the system as a result of this peak. The role of the vHs in particular in the chalcogenide spinels may be suitable for further study.

## 4. Conclusions

The $Cu(Ir_{1-x}Pt_x)_2Se_4$ (0 ≤ x ≤ 0.5) spinel has been synthesized via a conventional solid state reaction method. Characterization shows that Pt doping of the metallic non-superconducting $CuIr_2Se_4$ compound yields bulk BCS-like superconductivity for $Cu(Ir_{1-x}Pt_x)_2Se_4$ in the composition regime 0.1 ≤ x ≤ 0.35. $Cu(Ir_{0.8}Pt_{0.2})_2Se_4$ shows the maximum $T_c$ = 1.76 K and the highest measured electronic specific heat coefficient γ = 16.5 mJ mol$^{-1}$ K$^{-2}$. Increased Pt substitution decreases $T_c$ in the compositions that are metallic in the normal state, but eventually semiconducting behavior, with low temperature transport potentially dominated by 3D variable range hopping, is observed for $CuIrPtSe_4$. Electronic structure calculations show that the composition observed to display the highest $T_c$ is near an electron count where the calculated density of states is highest, and they also show $CuIrPtSe_4$ to be a conventional band-filling derived semiconductor. Although electron doping of $CuIr_2S_4$ by substitution of Zn for Cu results in superconductivity at a similar electron count (0.2 ≤ x ≤ 0.3 for $Cu_{1-x}Zn_xIr_2S_4$) to what is found in the current system, superconductivity cannot be induced by substitution of Pt for Ir, at least up to x = 0.3 in $Cu(Ir_{1-x}Pt_x)_2S_4$ [33]. The host compound for the current system, $CuIr_2Se_4$, does not display any of the complex electro-structural coupling phenomenology that has been observed for $CuIr_2S_4$, and therefore the appearance of superconductivity in $CuIr_2X_4$ chalcogenide spinels cannot strictly be associated with the suppression of that complex state. The occurrence of an M-I transition under pressure in CuIr$_2$Se$_4$ suggests that such instabilities are hidden just below the stability criterion in that compound. Thus if the instabilities have anything to do with the occurrence of superconductivity in the iridium chalcogenide spinels, it can only be that the tendency toward electro-structural instability is all that is required, not the actual physical manifestation of that instability through real structural or electronic phase transitions.

**Acknowledgements**

The calculations and low temperature specific heat measurements were supported by the department of energy, grant DOE FG02-98ER45706. The synthesis, X-ray, and other characterization were supported by the Air Force MURI on superconductivity, grant FAA9550-09-1-0593. TK thankfully acknowledges support from the Foundation for Polish Science (SKILLS - FNP Mentoring).

**Table 1 Superconducting and normal-state properties for $Cu(Ir_{1-x}Pt_x)_2Se_4$.**

|  | $Cu(Ir_{1-x}Pt_x)_2Se_4$ | | | | | |
|---|---|---|---|---|---|---|
|  | x = 0 | x = 0.1 | x = 0.2 | x = 0.3 | x = 0.4 | x = 0.5 |
| $T_C$ (K) | --- | 1.06 | 1.76 | 1.64 | --- | --- |
| a (Å) | 10.3199 | 10.3351 | 10.3453 | 10.3625 | 10.3742 | 10.3864 |
| $V_m$ (cm$^3$ mol$^{-1}$) | 82.73 | 83.10 | 83.35 | 83.76 | 84.05 | 84.34 |
| $\gamma$ (mJ mol$^{-1}$ K$^{-2}$) | 7.0(1) | 13.1(1) | 16.5(1) | 15.7(1) | 10.9(2) | 3.0(1) |
| $\Theta_D$ (K) | 222 | 225 | 212 | 227 | 211 | 249 |
| $\lambda_{ep}$ | --- | 0.51 | 0.57 | 0.56 | --- | --- |
| $N(E_F)$ experiment (states/eV/f.u.) | --- | 3.69 | 4.45 | 4.28 | --- | --- |
| $N(E_F)$ calculations (states/eV/f.u.) | 3.25 | 3.67 | 5.87 | 9.6 | 7.34 | 0 |
| $\Delta C/\gamma T_c$ | --- | --- | 1.58 | 1.44 | --- | --- |
| $H_{C2}(0)$ (KOe) | --- | --- | 32 | 36 | --- | --- |
| $\xi_0$ (Å) | --- | --- | 101 | 96 | --- | --- |

**References**

1. D.C. Johnston, J. Low Temp. Phys. **25,** 145 (1976).

2. R.W. Callum, D.C. Johnston, C.A. Luengo, and M.P. Maple, J. Low Temp. Phys. **25,** 177 (1976).

3. R. N. Shelton, D. C. Johnston, and H. Adrian, Solid State Commun. **20,** 1077 (1976).

4. M. Robbins, R. H. Willens, and R.C. Miller, Solid State Commun. **5,** 933 (1967).

5. T. Hagano, Y. Seki, N. Wada, S. Tsuji, T. Shirane, K.I. Kumagai, and S. Nagata, Phys Rev B **51,** 12673 (1995).

6. G. Cao, T. Furubayashi, H. Suzuki, H. Kitazawa, T. Matsumoto, and Y. Uwatoko, Phys. Rev. B **64**, 214514 (2001).

7. S. Nagata, T. Hagino, Y. Seki, and T. Bitoh, Physica B: Cond. Mat. **194,** 1077 (1994).

8. T. Furubayashi, T. Kosaka, J. Tang, T. Matsumoto, Y. Kato, and S. Nagata; J. Phys. Soc. Jpn. **66,** 1563 (1997).

9. G. Oomi, T. Kagayama, I. Yoshida, T. Hagino, and S. Nagata, J. Mag. Mag. Mat. **140,** 157 (1995).

10. K. Takubo, S. Hirata, J. Y. Son, J. W. Quilty, T. Mizokawa, N. Matsumoto, and S. Nagata, Phys. Rev. Lett. **95,** 246401 (2005).

11. P. G.Radaelli, Y. Horibe, M.J. gutmann, H. Ishibashi, C. H. Chen, R. M. Ibberson, Y. Koyama, Y.S. Hor, V. Kiryukhin, and S. W. Cheong, Nature **416,** 155 (2002).

12. P. G. Radaelli, New J. Phys. **7,** 53 (2005).

13. E. Morosan, H. W. Zandbergen, B. S. Dennis, J. W. G. Bos, Y. Onose, T. Klimczuk, A. P. Ramirez, N. P. Ong, and R. J. Cava, Nat. Phys. **2,** 544 (2006).

14. A.W. Sleight, J. L. Gillson, and P. E. Bierstedt, Solid State Commun. **17,** 27 (1975).

15. C. Niedermayer, C. Bernhard, C. Blasius, T. Golnik, A. Golnik, A. Moodenbaugh, and J.I. Budnick, Phys. Rev. Lett. **80,** 3843 (1998).

16. M. D. Lumsden, and A. D. Christianson, J. Phys. Cond. Matt. **22,** 203203 (2010).

17. S. Nagata, N. Matsumoto, Y. Kato, T. Furubayashi, T. Matsumoto, J. P. Sanchez, and P. Vulliet, Phys. Rev. B **58,** 6844 (1998).

18. A. T. Burkov, T. Nakama, K. Shintani, K. Yagasaki, N. Matsumoto, and S. Nagata, Physica B: Cond. Mat. **281,** 629 (2000).

19. T. Hagino, Y. Seki, and S. Nagata, Physica C: Superconductivity **235,** 1303 (1994).

20. S. Nagata, N. Matsumoto, R. Endoh, and N. Wada, Physica B: Cond. Matt. **329,** 944 (2003).

21. L. Chen, M. Matsunami, T. Nanba, T. Matsumoto, S. Nagata, and Y. Ikemoto, J. Phys. Soc. Jpn. **74,** 1099 (2005).

22. A. T. Burkov, T. Nakama, M. Hedo, K. Shintani, K. Yagasaki, N. Matsumoto, and S. Nagata, Phys. Rev. B **61,** 10049 (2001).

23. P. Blaha, K. Schwarz, G.Madsen, D. Kvasnicka and J. Luitz, WIEN2k, An Augmented Plane Wave + Local Orbitals Program for calculating Crystal Properties, Technische Universitat Wien, Austria, 2001.

24. J. P. Perdew, K. Burke, and M. Ernzerhof, Phys. Rev. Lett. **77,** 3865 (1996).

25. T. Klimczuk, and R. J. Cava, Phys. Rev. B **70,** 212514 (2004).

26. N. R. Werthamer, E. Helfand, and P. C. Hohenberg, Phys. Rev. **147,** 295 (1966).

27. N.F. Mott, Phil. Mag. **19,** 835 (1969).

28. W. L. McMillan, Phys. Rev. **167**, 331 (1968).

29. J.E. Hirsch and D.J. Scalapino, Phys. Rev. Lett. **56**, 2732-2735 (1986).

30. I.E. Dzaloshinskii, Zh. Eksp. Teor. Fiz. **93**, 1487 (1987).

31. C. Felser, K. Thieme, and R. Seshadri, J. Mat. Chem. **9,** 451 (1999).

32. C. Felser, J. Alloys Compd. **262,** 87 (1997).

33. N. Matsumoto, Y. Yamauchi, J. Awaka, Y. Kamei, H. Takano and S. Nagata, Int. J. Inorg. Mat. **3,** 791 (2001).

**Figures**

**Figure 1** (Color online) (a) Composition dependence of the room temperature lattice parameter for $Cu(Ir_{1-x}Pt_x)_2Se_4$ (0.0 ≤ x ≤ 0.5). Inset: The spinel crystal structure: CuSe4 tetrahedra, blue; (Ir,Pt) octahedra, gray; Se ions green. (b) XRD patterns of $Cu(Ir_{1-x}Pt_x)_2Se_4$ (0.1 ≤ x ≤ 0.5) compounds heated at 850 °C for 96 h and $Cu(Ir_{0.5}Pt_{0.5})_2Se_4$ heated for 96 + 48 + 72h.

**Figure 2** (Color online) The temperature dependence of the electrical resistivity of polycrystalline samples of the $Cu(Ir_{1-x}Pt_x)_2Se_4$ (0.05 ≤ x ≤ 0.4) compounds without magnetic field; Inset: Enlarged view of low temperature region (0.4 - 3 K) of the electrical resistivity of $Cu(Ir_{1-x}Pt_x)_2Se_4$ (0.1 ≤ x ≤ 0.35).

**Figure 3** (Color online) Low temperature resistivity at various applied fields for (a) $Cu(Ir_{0.8}Pt_{0.2})_2Se_4$ and (b) $Cu(Ir_{0.7}Pt_{0.3})_2Se_4$. Inset shows the temperature dependence of the upper field ($H_{c2}$).

**Figure 4** (a) Temperature dependence of the electrical resistivity of polycrystalline sample of $CuIrPtSe_4$; inset: Temperature dependence of the Seebeck coefficient for $CuIrPtSe_4$. (b) Temperature dependence of the resistivity as log ρ vs. (1/T). Inset: low temperature data plotted as log ρ vs. $(1/T)^{1/4}$.

**Figure 5** (Color online) Temperature dependence of specific heat $C_p$ of $Cu(Ir_{0.8}Pt_{0.2})_2Se_4$ measured under magnetic fields 0 T and 5 T, presented in the form of $C_{el}/T$ vs T (main panel) and $C_p/T$ vs $T^2$ (inset). The fitting of the low temperature with the range 2 - 7 K heat capacity data obtained in the magnetic field 5 T.

**Figure 6** (Color online) (a) The Pt content dependence of electronic specific-heat coefficients (γ) and (b) the Debye temperature ($\Theta_D$) obtained from low temperature fits of specific heats. (c) The electronic phase diagram for $Cu(Ir_{1-x}Pt_x)_2Se_4$ (0 ≤ x ≤ 0.5) as a function of Pt content x.

**Figure 7**.(Color online) (a) The calculated Density of states and the Fermi surface for Cu(Ir$_{0.7}$Pt$_{0.3}$)$_2$Se$_4$. The inset shows the dependence of the calculated DOS at E$_F$ on the doping level. Two different projections of the Fermi surface are shown. The colors are a guide for the eye to emphasize the topography. (b) The band structure close to E$_F$ of Cu(Ir$_{0.7}$Pt$_{0.3}$)$_2$Se$_4$. A vHs with a very flat band is visible at the Γ-point.

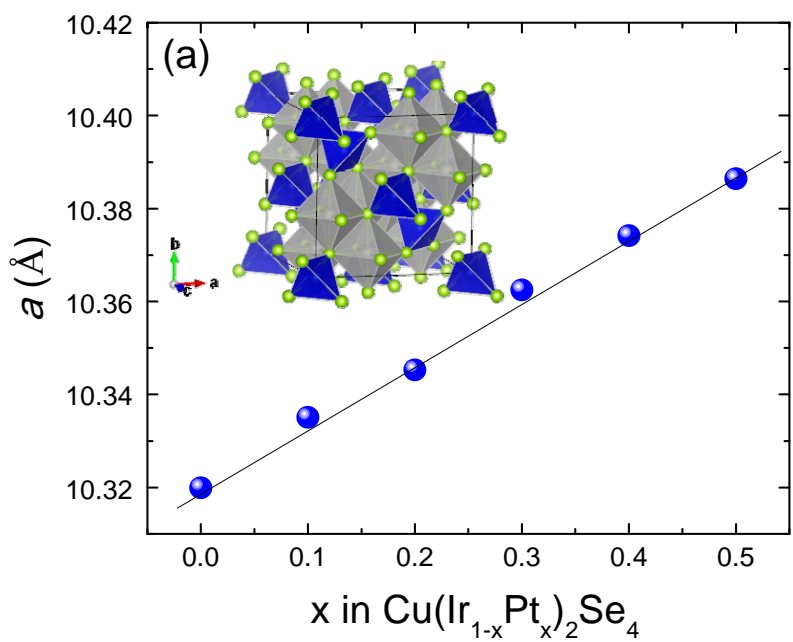

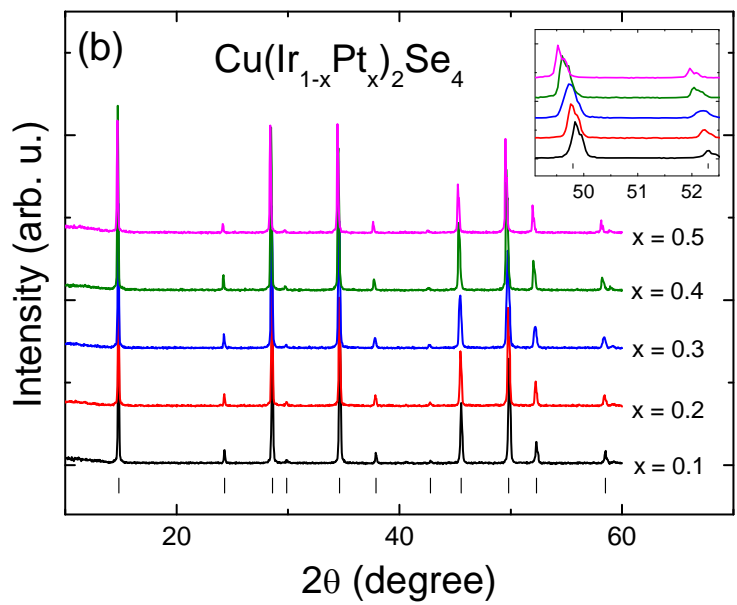

**Fig. 1 (a) and (b)**

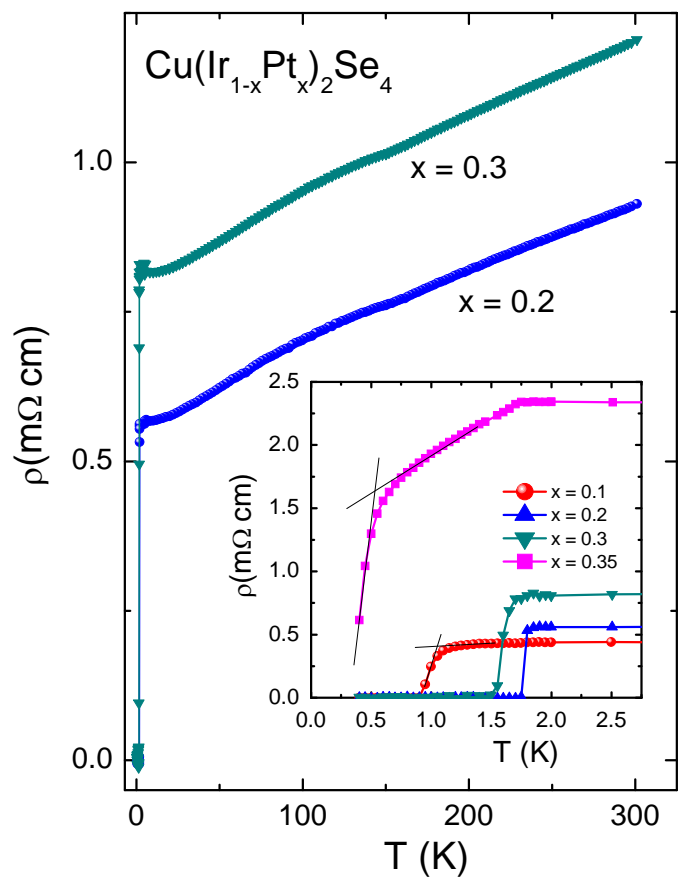

**Fig. 2**

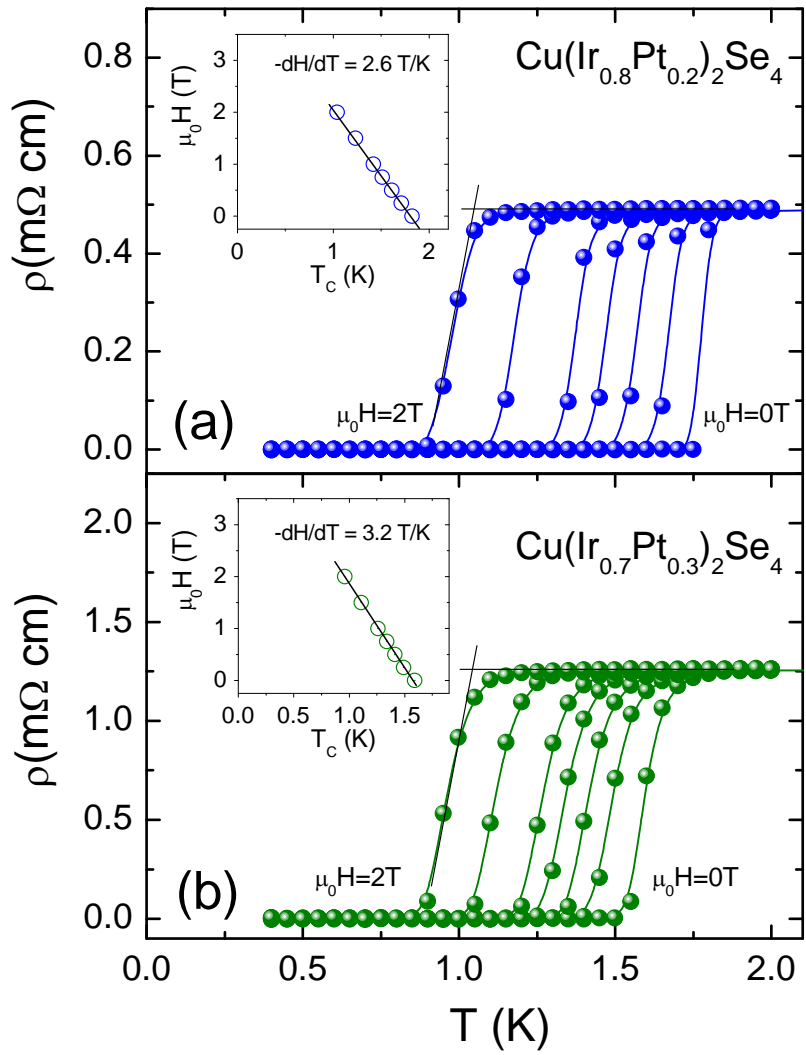

**Fig. 3(a) and (b)**

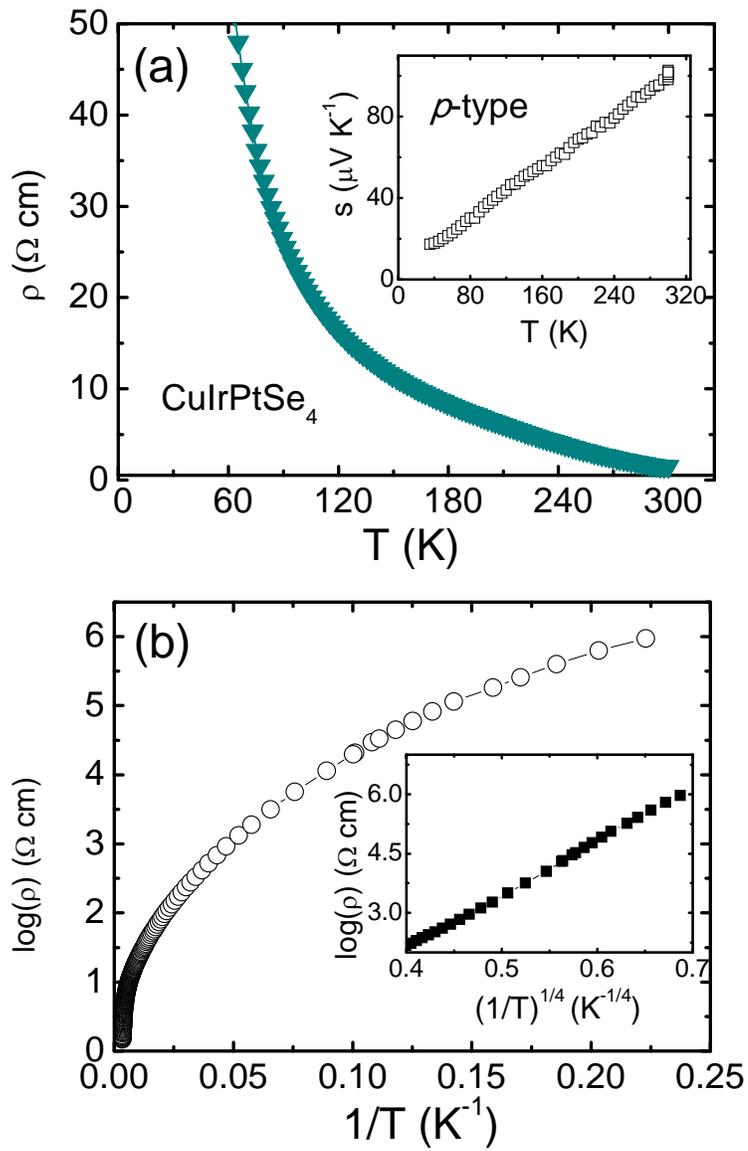

**Fig. 4(a), (b)**

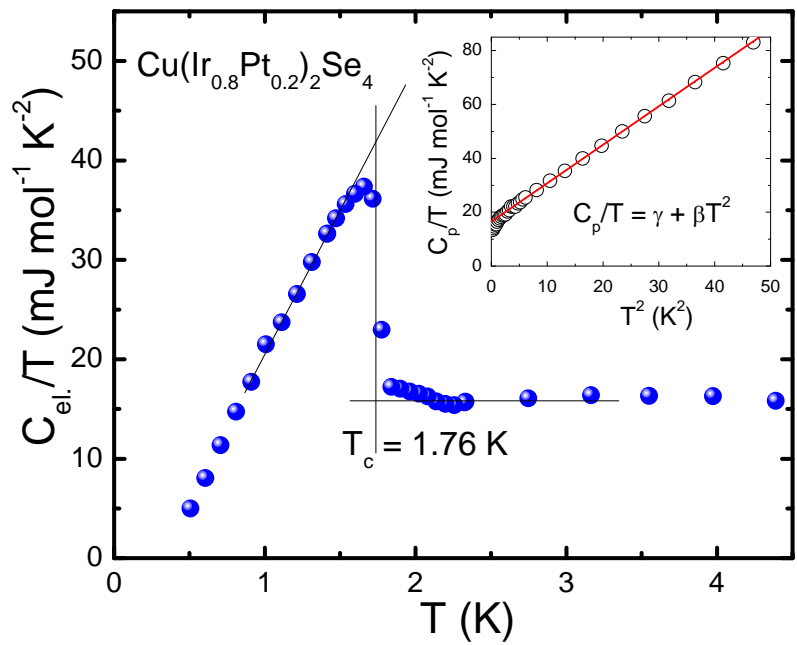

**Fig. 5**

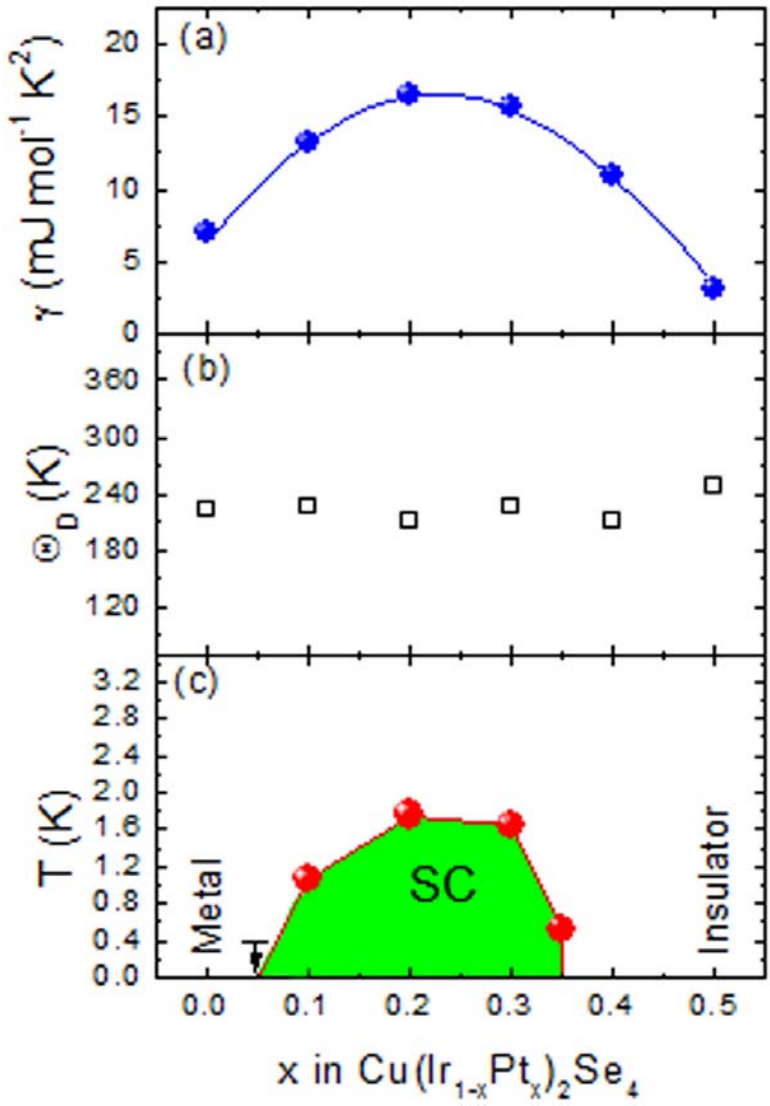

**Fig. 6(a), (b) and (c)**

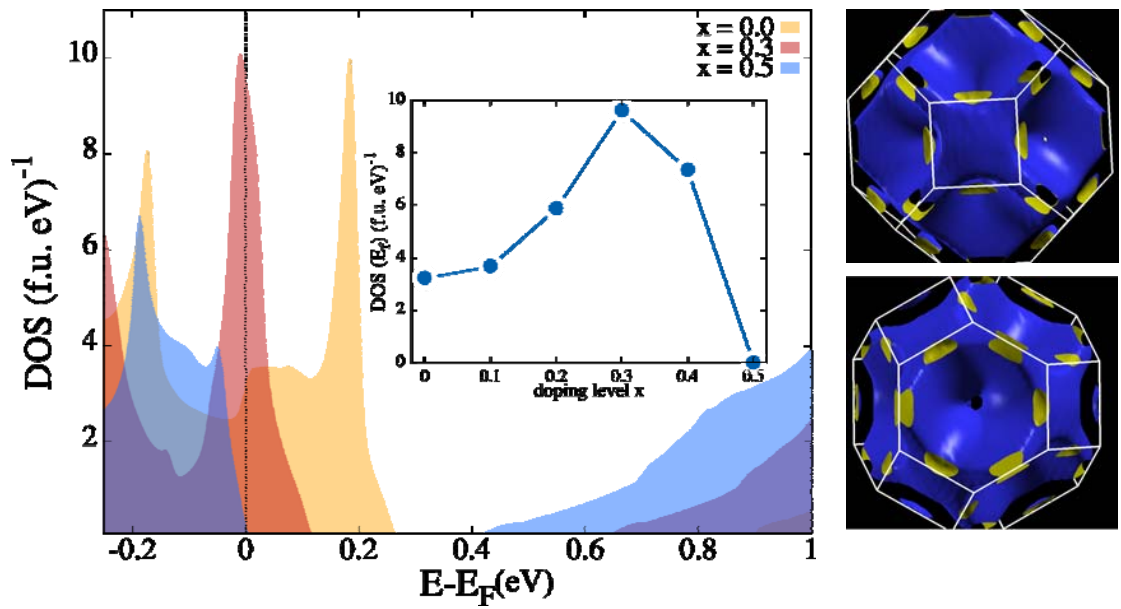

**Fig. 7(a)**

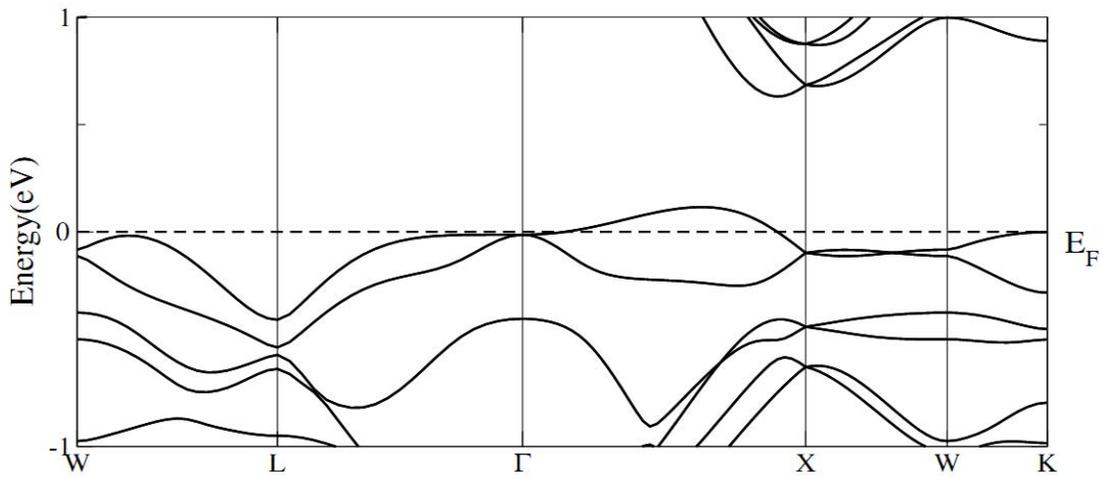

**Fig. 7(b)**